\documentstyle[12pt]{article}

\newcommand{\Section}[1]{\section{#1}\setcounter{equation}{0}}

      \def\be{\begin{equation}}
      \def\bea{\begin{eqnarray}}
      \def\ba{\begin{array}}
      \def\ea{\end{array}}
      \def\poi{Poincar\'e \,}
      \def\lor{Lorentz \,}
      \def\rot{SO(3) {\rm -like}}
      \def\pot{SO(2,1) {\rm -like}}
      \def\eot{E(2) {\rm -like}}
      \def\eijk{\epsilon_{ijk}}
      \def\sh0{\sinh({P_{0} \over \kappa})\,}
      \def\s0{\sin({P_{0} \over \kappa})\,}
      \def\ch0{\cosh({P_{0} \over \kappa})\,}
      \def\c0{\cos({P_{0} \over \kappa})\,}
      \def\k{\kappa}
      \def\ee{\end{equation}}
      \def\eea{\end{eqnarray}}
      \def\R{\rm {I\kern-.200em R}}
      \def\C{\rm {I\kern-.520em C}}
      \def\pex{\exp({{i P_0} \over {2 \k}})}
      \def\mex{\exp(-{{i P_0} \over {2 \k}})}

\begin{document}
\vskip 50mm
{\large
   {\bf
       \centerline{ On Little Groups and Boosts of }
       \centerline{$\kappa$-deformed \poi Group }
         } 
 } 
\vskip 10 mm
\centerline{ S. Rouhani and A. Shariati }
\vskip 10 mm
{\it
  \centerline{
             Department of Physics, Sharif University of Technology,
             }
  \centerline{
             P.O.Box 9161, Tehran 11365, Iran.
             }
  \centerline{
              And
              }
  \centerline{
             Institute for Studies in Theoretical Physics and
              Mathematics,
             }
  \centerline{
             P.O.Box  5746, Tehran 19395, Iran.
             }
    } 
\vskip 10 mm

\begin{abstract}
 We show how Wigner's  little group  approach  to the  representation theory
 of  Poincar\'e  group may be generalized  to the case of  $\kappa$-deformed
 Poincar\'e group.  We also derive  the deformed  Lorentz transformations of
 energy and momentum. We find that if the $\kappa$-deformed Poincar\'e group
 is adopted as the fundamental symmetry of nature,  it results in deviations
 from predictions of the Poincar\'e symmetry at large energies, which may be
 experimentally observable. \end{abstract}

\newpage

\Section{
          Introduction
          }

Since the discovery of quantum groups as generalizations of ordinary
groups, there is a tendency in theoretical physics to apply these objects
in physical theories. In high energy physics groups of symmetries appear
at various levels. First, there is the group of symmetries of the space-time
manifold. In the standard model this is the \poi group. Second, there are the
gauge groups : $SU(3)$, $SU(2)\times U(1)$, or other groups in the grand
unified theories. There also appear some groups related to flavor.
Any of these groups may be deformed, here we only consider the deformation
of the \poi group.
We shall not go into the details of the idea that lies behind the quantization
of groups. We only mention that although quantum groups were first introduced
in the context of integrable models, nowadays people try
to use them in other problems, hoping that a quantum group may be
a better tool to formulate the symmetry of a theory.
Our interest is the structure of space-time, therefore
we consider the \poi group.
This group is related to the structure of space-time and we know,
from Einstein's general relativity, that gravity affects this structure.

Being the semidirect product of \lor and translation groups,
the \poi group is not semisimple, therefore, its quantization is not unique.
Up to now there have been discovered three different deformations of
\poi group \cite{LNRT,OSWZ,SWW}.
In this paper we consider $\k$--deformations of the \poi group as introduced
by Lukierski et al. \cite{LNRT}.
We prefer this deformation because it is minimal in the sense that only
two of the commutation relations are
deformed and there is a very clear correspondence between the generators
of this quantum group and the generators of the un-deformed \poi group.
Furthermore, there are some physical reasons in choosing this deformation
\cite{Bacry}.

Our aim in this paper is to get some insight in the consequences of this
deformation from
a physical point of view and obtain some of the deviations that these
deformations imply up to first order in observable quantities such as
energy and momenta.
Whether or not these
deviations may be tested experimentally requires more careful study.
Specifically, one has to take into account any other perturbative effect that
has the same order.

In section 2 we briefly review the \poi group and its $\k$-deformed version.
In section 3 we apply Wigner's little group approach to the $\k$-\poi group.
In section 4 we see that the non-trivial co-product of $\k$-\poi leads to a
             feature that is not present in the \poi group.
In section 5 we consider boosts, infinitesimal and finite.
We end this paper by conclusion in section 6.

\Section{
         The \poi group and its $\k$--deformations
         }

The \poi group is generated by 10 generators: $P_{0}$, $P_{i}$, $M_{i}$,
$L_{i}$
for $i=1,2,3$. The commutation relations are:
\be
 \label{comm}
 \ba{lll}
   [P_{0},P_{i}]=0                      &
   [P_{i},P_{j}]=0                      &
   [M_{i},M_{j}]= i\eijk M_{k}          \cr
   [M_{i},P_{0}]=0                      &
   [M_{i},P_{j}]= i\eijk P_{k}          &
   [M_{i},L_{j}]= i\eijk L_{k}          \cr
   [L_{i},P_{0}] = i P_{i}              &
   [L_{i},P_{j}] = i \delta_{ij} P_{0}  &
   [L_{i},L_{j}] = -i \eijk M_{k}       \cr
 \ea
\ee
The $\k$--deformation is obtained by a contraction of $U_{q}(SO(2,3))$ which
is the deformed anti de Sitter algebra \cite{LNRT}.
The commutation relations for the
$\k$--\poi algebra are exactly the same as those for the \poi algebra except
for the following two relations:
\be
 \label{nine}
 \ba{l}
  [L_{i},P_{j}] = i \delta_{ij} \k \sh0     \cr
  [L_{i},L_{j}] = -i \eijk (M_{k} \ch0 - {1 \over {\k^2}} P_{k} P_{l} M_{l}) .
 \ea
\ee
Here $\k$ is a constant with the dimensions of energy. Usually $\k$ is
regarded as real, however, a pure imaginary one also leads to a Hopf algebra.

The deformed Pauli--Lubanski four-vector is defined as:
\be
 \ba{ll}
   W_{0} = P_{i} M_{i}  &
   W_{i}= \k \sh0 M_{i} + \eijk P_{j} L_{k} .
 \ea
\ee
The two Casimir operators are
\be
 \label{twelve}
 \ba{l}
  c_{1} = 4 \k^2 \sinh^{2}({ P_{0} \over { 2 \k}}) - P_{i} P_{i}   \cr
  c_{2} = ( \ch0 - {{1} \over {4 \k^2}} P_{i} P_{i} ) W_{0}^2 -
  W_{i} W_{i} .
 \ea
\ee
We note that the $\k \rightarrow \infty $ limit of the above relations
are the familiar ones for the \poi algebra. Therefore, if $\k$--\poi is the
correct symmetry algebra of nature $\k $ must be large. If we want to have
real energy $P_{0} $ and momenta $P_{i} $ then $\k $ must be real or pure
imaginary. Imaginary $\k $ has the odd property that the total energy and
total momenta of a system composed of two subsystems will not be real : the
co-product of the corresponding Hopf algebra contains $\exp(i{P_{0} \over
\k})$,
cf. \cite{GKMMK}. For completeness we write the co-product and antipode.
\be
 \ba{l}
  \Delta(P_0) = P_0 \otimes 1 + 1 \otimes P_0         \cr
  \Delta(M_i) = M_i \otimes 1 + 1 \otimes M_i         \cr
  \Delta(P_i) = P_i \otimes \pex + \mex \otimes P_i   \cr
  \Delta(L_i) = L_i \otimes \pex + \mex \otimes L_i   \cr
  \hskip 17 mm  + {i \over {2 \k}} \epsilon_{ikl}
                \left( \pex M_k \otimes P_l + P_k \otimes M_l \pex \right)
 \ea
\ee
\be
 \ba{lll}
  {\rm S}(P_{\mu}) = - P_{\mu}                &
  {\rm S}(M_i) = - M_i                        &
  {\rm S}(L_i) = L_i - {3 \over {2 \k}} P_i
 \ea
\ee

\Section{
         Mass-shells and little groups
         }

The idea of using little groups to obtain information about the representations
of the \poi group is due to Wigner \cite{Wigner}. For a complete
accounting of this we refer the reader to \cite{KN}.
Here we briefly review this idea and try to apply it to the $\k$-deformed
\poi group.

Because of (\ref{comm}), we can diagonalize energy and momenta simultaneously
and consider physical states with definite energy and momenta
\( \vert P_0,P_i > \).  Next we look at at the orbits of these states and
the little groups.
By an orbit we mean a $c_{1} = {\rm constant} $
surface in the $\R^{4} \sim \{ (P_{0},P_1,P_2,P_3) \} $.
For the undeformed \poi group these orbits are : light-cone for a
massless particle, two time-like hyperbolas for a massive particle and a
hyperboloid of revolution for a tachyon. A general (finite) action of the \poi
group can change the four-momentum of a particle but only on its mass-shell.
We know also that the action of the \lor group on these orbits is
transitive\footnote{
                     We omit the origin $P_{\mu} =0$ from the light-cone.
                      }.

Although for the $\k $--deformations the finite action is not well-understood,
we know that the orbits; i.e. the mass shells; are stable. This is because
$c_{1}$ is the Casimir operator.
However we don't know about the transitivity of this action.

For a given $ P_{\mu} $ the little group is
defined as the subgroup of the \lor group that leaves the given $ P_{\mu} $
intact. The Lie algebra of this little group is taken as the subset of
the algebra defined by eq (\ref{comm}) which is closed and leaves a physical
state invariant. This Lie algebra is the Lie algebra of $SO(3)$,
$E(2)$ or $SO(2,1)$ for time-like, light-like and space-like four-momenta
respectively.  The topology of these Lie groups may be quite different from
$SO(3)$, $E(2)$ and $SO(2,1)$, therefore we call them $\rot$, $\eot$ and
$\pot$\footnote{
                For example, the little group of electrons is $SU(2)$
                which is the double cover of $SO(3)$.
               }.
Let's take a look at the orbits.
For the undeformed \poi group these are the well-known hyperboloids
$ E^2 - P^2 = c_1 $
where $E$ is for $P_{0} $, $P$ is
for $P_{3}$ and $P_{1} = P_{2} = 0$.
For the real $\k$-deformation the mass-shells are topologically like
the ones in special relativity. The defining equation in this case is
$ 4 \k^2 \sinh({E \over {2 \k}}) - P^2 = c_1 $.
For the imaginary $\k $ this equation becomes
$ c_{1} = 4 \chi^2 \sin^{2}({ E \over { 2 \chi}}) - P^2 $ where $\k = i \chi $.
For this $\chi$-deformation we note that:
(i) Because of the periodicity of $\sin^{2}({ E \over { 2 \chi}})$ the levels
$E = 2 n \pi \chi $ are identified, therefore, all orbits are closed.
(ii) For a particle with definite $c_{1}$ there is a bound for momentum.
If $c_{1} < 0 $; i.e. for tachyons; there is a lower and an upper
bound for momentum and no bound for energy except the periodicity of energy.
If $ c_{1} \geq 0 $ there is only an upper bound for the
momentum and a lower bound for the energy: the rest mass. For these massive
particles there is an upper bound for $ c_{1} $ itself $ c_{1} < 4
\chi^2 $. A particle with the critical value $ c_{1} = 4 \chi^2 $ has zero
momentum in any frame.
Particles with mass near this critical value
has almost zero momenta in all frames.
This has no correspondence in special relativity; i.e. in \poi group; or
in the real $\k $--deformation.
In special relativity massless
particles in all frames have the same velocity but their energy and
momenta differ. Now in a $\chi$--deformed special relativity we face with
the possibility of the existence of particles with the property that
in all frames they have the same energy and zero momenta. It is
interesting to study the velocity of these particles in different frames.

Wigner's approach to the representation theory of the \poi group is based
on the idea of little groups.
Since for the deformed \poi group the momenta commute; $[P_{\mu},P_{\nu}]=0$;
this idea may be applied to them also. The key idea
is that:
\be
  \label{fouteen}
  [W_{\mu},P_{\nu}] = 0
\ee
so that the the action of the $ W_{\mu} $ leaves $ P_{\nu} $ intact. The
remaining point is whether $ \{ W_{\mu} \}_{\mu = 0}^{3} $ generate a
subgroup of the \lor group.
After a little calculations one can prove that:
\be
 \ba{l}
  [ W_{0},W_{1}] = i ( W_{2} P_{3} - W_{3} P_{2} )    \cr
  [ W_{1},W_{2}] = i ( W_{1} \alpha P_{1} P_{3} + W_{2} \alpha P_{2} P_{3}
                      + W_{3} (\alpha P_{3}^{2} + \xi) )
 \ea
\ee
and relations obtained from these by cyclic permutations of the indices. Here
we have used the notations:
\be
  \xi = \k \sh0
  \qquad
  \alpha \, \xi = {1 \over \k^2} P_{i} P_{i} - \ch0 .
\ee
Note that in all these commutation relations momenta have gone to the right.
This means that acting on eigenvectors of the momenta they are c-numbers,
therefore, they can be moved to the left and all the commutation relations are
of the following form:
\be
  [W_{\mu},W_{\nu}] = C^{\sigma}_{\mu \nu} W_{\sigma}
\ee
where $ C^{\sigma}_{\mu \nu} $ s are the structural constants depending on
the given four-momentum.
So, the little groups of the $\k$-- and $\chi$--deformations of the
\poi group are ordinary Lie groups.
Because there is a linear relation among the $ W $s,
\be
  \xi W_{0} - P_{i} W_{i} = 0,
\ee
this Lie algebra is three dimensional. To reveal its
structure we consider cases such that:
\be
  P_{0} = E \qquad P_{1} = 0 \qquad P_{2} = 0 \qquad P_{3} = P .
\ee
We get:
\be
  W_{0} = P M_{3}
  \qquad
  W_{1} = \xi M_{1} - P L_{2}
  \qquad
  W_{2} = \xi M_{2} + P L_{1}
  \qquad
  W_{3} = \xi M_{3}
\ee
where $ \xi = \k \, \sinh({E \over \k}) $.
Note that $ W_{0} $ and $ W_{3} $ are proportional and for $ E \rightarrow 0 $,
$ W_{0} $ is well-defined and $ W_{3} = 0 $. For the Lie algebra we get:
\be
  [W_{1},W_{2}] = i \, \xi \, \, f(E,P) \, W_{3}
  \qquad
  [W_{2},W_{3}] = i \, \xi \, W_{1}
  \qquad
  [W_{3},W_{1}] = i \, \xi \, W_{2}
\ee
where
\be
  \label{ffunction}
  f(E,P) = \k^2 \sinh^2({E \over \k}) - P^2 ( \cosh({E \over \k}) -{P^2 \over
          4 \k^2} )  = c_{1} ( 1 + { c_{1} \over 4 \k^2} ) .
\ee
{}From these we conclude that the little groups for the $\k$--deformation are:
\be
  \ba{lllll}
    a & \rot & {\rm for} & c_{1} > 0             & {\rm massive}       \cr
    b & \eot & {\rm for} & c_{1} = 0             & {\rm massless}      \cr
    c & \pot & {\rm for} & - 4 \k^2 < c_{1} < 0  & {\rm tachyonic\, A} \cr
    d & \eot & {\rm for} & c_{1} = -4 \k^2       & {\rm tachyonic\, B} \cr
    e & \rot & {\rm for} &  c_{1} < - 4 \k^2     & {\rm tachyonic\, C} \cr
  \ea
\ee
\vskip 5 mm
Cases $d$ and $e$ have no counterpart in the \poi group. For the
$\chi$--deformation the little groups are:
\be
  \ba{lllll}
    a' & \rot & {\rm for} & 0 < c_{1} < 4 \chi^2 & {\rm massive}     \cr
    b' & \eot & {\rm for} & c_{1} = 0            & {\rm  massless}   \cr
    c' & \pot & {\rm for} & c_{1} < 0            & {\rm tachyonic}   \cr
    d' & \eot & {\rm for} & c_{1} = 4 \chi^2     & {\rm massive}     \cr
  \ea
\ee
\vskip 5 mm
Case $d'$ has no counterpart in the \poi group.
Therefore, depending on the values for $P_{0}$,$P_{1}$,$P_{2}$, and $P_{3}$,
we get different little groups.
In the $\k$--deformation there is a
region which is not present in the undeformed case where for
a space like four-momentum, i.e. a tachyon, the little group  is always
$\pot$. In
the $\k$--deformed \poi group tachyons divide into three classes. In one
class, which we name type A,
near the light cone, the little group is still $\pot$. For the special value
$c_{1} = -4 \k^2$ the little group is $\eot$ and we name such a tachyon of
type B. For type C we have
$c_{1} < - 4 \k^2$, the little group is $\rot$.
In the imaginary $\k$--deformed
\poi group there is a time like orbit consisting of just one point: $E=  \pi
\chi$, $P_{1}=P_{2}=P_{3}=0$ . For this orbit the little group is $\eot$.
The little group for the vacuum, i.e. $P_{\mu}=0$, is $SO(3,1) {\rm -like}$ in
all three cases.

It is interesting that the function $f$ in (\ref{ffunction}) is the
Casimir operator found recently by Ruegg et al. in \cite{RT}. According to
their results this is the ``covariant'' length squared of the four-momentum:
$P_{\mu} P^{\mu}$.

\Section{
         Little group of a composite system
         }

Ordinarily one can consider a system composed of two non-interacting particles.
This composite system must carry a representation of the \poi group. One
can therefore speak about its little group. In the (non-deformed) \poi
group, if the two particles have identical energy and momenta; i.e. they
have identical rest mass and are at rest with respect to each other;
then, the composite system has the same little group type.
For example, the little group of two electrons is $\rot$. In the $\chi$-- and
$\k$--deformations there are some differences.
In the $\k$--deformation the composite system made up of two tachyons
of type A may be of type B or C; and a system composed of two tachyons of
type B is of type C. To see this we use the co-product
to obtain the structural constants of the composite system.
Since the two particles have the same energy and momenta, and because of
the form of the co-product for $\k$--\poi group, we have
\be
  E_{{\rm tot}} = 2 \, E
  \qquad
  P_{{\rm tot}} = 2 \,  P \, \cosh({E \over {2 \k}})
\ee
Therefore, we obtain:
\be
  f_{{\rm tot}} = f\left( 2 E , 2 P \cosh({E \over {2 \k}})\right)
                = 4 \cosh^{2}({E \over {2 \k}}) \, c_{1} \,
                  \left( \k^2 + \cosh^{2}({E \over {2 \k}}) \, c_{1}
                                                             \right) .
\ee
{}From this last equation we see that if $c_{1}$ is negative then the sign of
$f_{{\rm  tot}}$ may be different from the sign of $f$. Setting $\chi = -i \k$
we see that this may happen for positive $c_{1}$ of $\chi$--deformation, where
the little group of two ordinary particles may be $\eot$.

\Section{
         Deformation of the boost operators
          }

The \poi group is the symmetry group of particles in the  relativistic quantum
mechanics. There must be a representation of the \poi group acting on the
Hilbert space of the particle.
In the rest frame of a particle there is a non-trivial
part of the \poi group that leaves the particle at rest, viz. the little
group. However, since the particle is at rest, the angular momentum generators
must act on the spin degrees of freedom. So we have
\be
  J_{i} = M_{i} + S_{i} \qquad B_{i} = L_{i} + K_{i}
\ee
where $J$ is the total angular momentum, $M$ is the orbital angular momentum,
$S$ is the spin, $B$ is the total boost operator, $L$ is the operator that
boosts the space-time coordinates and  $K$ is the operator that boosts the
spin degrees of freedom. This form of total angular momentum and boost come
from
the trivial co-product of the \poi group:
\be
  \Delta(J_{i}) = J_{i} \otimes 1 + 1 \otimes J_{i} = M_{i} + S_{i}
  \qquad
  {\rm etc.}
\ee
For the $\k$--deformation we use the non-trivial co-product to obtain:
\be
  \label{ntcp}
  J_{i} = M_{i} + S_{i}
  \qquad
  B_{i} = L_{i} + \exp(- { P_{0} \over {2 \k} } ) K_{i} + { 1 \over {2 \k} }
  \eijk P_{j} S_{k}.
\ee
$ \{ S_{i}, K_{i} ,i=1,2,3 \}$ generates a non-deformed \lor group. The new
feature of the $\k$--deformed \poi group is (\ref{ntcp}) which states that the
effect of a boost depends on the energy-momentum four-vector. Expanding
(\ref{ntcp}) to first order in $ 1 \over \k$ we get:
\be
  B_{i}=L_{i}+(1-{P_{0} \over{2 \k}})K_{i}-{1 \over{2 \k}} \eijk P_{j}
        S_{k}.
\ee
For our  standard case, i.e. $P_{0}=E$,$P_{1}=P_{2}=0$ and $P_{3}=P$ we get:
\be
 \label{tboosts}
 \ba{l}
  B_{1} = L_{1} +(1-{E \over{2 \k}})K_{1} - {1 \over{2 \k}} p S_{2}  \cr
  B_{2} = L_{2} +(1-{E \over{2 \k}})K_{2} + {1 \over{2 \k}} p S_{1}  \cr
  B_{1} = L_{3} +(1-{E \over{2 \k}})K_{3}                            \cr
 \ea
\ee

Now we turn to finite \lor (boost) transformations.
A glance at the commutation relations  shows that the finite action of the
rotation subgroup is not deformed. So is the action of translations on the
angular momenta. The deformation appear only in boosts. Let's consider the
effects of a finite boost in the $z$ direction on $E = P_{0}$ and $P = P_{3}$.
\be
  \label{eprime}
  E \longrightarrow E' = \exp( -i \eta L) E \exp(i \eta L) =
\sum_{n=0}^{\infty}
                         {(i \eta)^n \over n!} L^{n}(E)
\ee
\be
  \label{pprime}
  P \longrightarrow P' = \exp( -i \eta L) P \exp(i \eta L) =
\sum_{n=0}^{\infty}
                         {(i \eta)^n \over n!} L^{n}(P)
\ee
where $\eta$ is the rapidity; $L = L_{3}$ and
\be
  L^{0}(x) = x \qquad L^{n+1}(x) = [ L , L^{n}(x) ].
\ee
In writing (\ref{eprime}) and (\ref{pprime}) we have assumed that the
action of a generator $X$ on an operator $\Omega$ is given by
\be
  \delta(\Omega) = i [ X , \Omega ]
\ee
as usual.

For the \poi group (\ref{eprime}) and (\ref{pprime}) lead to
the familiar \lor
transformations of energy and momentum. We want to see the effect of
deformations on these transformations. Using commutation relations one can
calculate $E'$ and $P'$ to any order $n$ in $\eta$. The result will be two
polynomials in $\sinh({E \over \k})$, $\cosh({E \over \k})$ and $P$. In this
form the approximation is in ignoring $O(\eta^{n+1})$ and everything is exact
in $\k$. Let's try to calculate everything to first non-zero order in
${1 \over \k}$ but to all orders in $\eta$. In this form the deformation is
seen more transparently. To this end we write everything up to first order in
${1 \over {6 \k^2}}$.
\be
 \ba{l}
  L(E)  = i P \cr
  L(P)  =  i ( E + {1 \over {6 \k^2}} E^{3} ) \cr
  L({1 \over {6 \k^2}} E^3)  =  i  {3 \over {6 \k^2}} E^{2} P \cr
  L({1 \over {6 \k^2}} E^2 P)  =  i {1 \over {6 \k^2}} (2 E P^2 +E^3) \cr
  L({1 \over {6 \k^2}} E P^2)  =  i {1 \over {6 \k^2}}(P^3 + 2 E^2 P) \cr
  L({1 \over {6 \k^2}} P^3)  =  i {3 \over {6 \k^2}}  E P^2 \cr
 \ea
\ee
In matrix form this can be written as
\be
  \label{action}
  \pmatrix{
            E                          \cr
            P                          \cr
           {1 \over  {6 \k^2}} E^3     \cr
           {1 \over  {6 \k^2}} E^2 P   \cr
           {1 \over  {6 \k^2}} E P^2   \cr
           {1 \over  {6 \k^2}} P^3     \cr
            }
  \longrightarrow
  \pmatrix{
            E'                             \cr
            P'                             \cr
           {1 \over  {6 \k^2}} {E'}^3      \cr
           {1 \over  {6 \k^2}} {E'}^2 P    \cr
           {1 \over  {6 \k^2}} E' {P'}^2   \cr
           {1 \over  {6 \k^2}} {P'}^3      \cr
            }
  =
  \pmatrix{
           0  &   1   &   0   &   0   &   0   &   0  \cr
           1  &   0   &   1   &   0   &   0   &   0  \cr
           0  &   0   &   0   &   3   &   0   &   0  \cr
           0  &   0   &   1   &   0   &   2   &   0  \cr
           0  &   0   &   0   &   2   &   0   &   1  \cr
           0  &   0   &   0   &   0   &   3   &   0  \cr
            }
  \pmatrix{
            E                          \cr
            P                          \cr
           {1 \over  {6 \k^2}} E^3     \cr
           {1 \over  {6 \k^2}} E^2 P   \cr
           {1 \over  {6 \k^2}} E P^2   \cr
           {1 \over  {6 \k^2}} P^3     \cr
            }
\ee
Now to calculate the effect of a finite boost with rapidity $\eta$ to
order ${ 1 \over {6 \k^2}}$ one has to compute $\exp(-\eta \Lambda)$ where
$\Lambda$ is the $6 \times  6$ matrix in (\ref{action}). The
resulting matrix gives the transformed energy $E'$ and momentum $P'$.
Because of the form of $\Lambda$ the transformation has the following form:
\be
  E' = \cosh(\eta) E - \sinh(\eta)P + {1 \over {6 \k^2}} \left(a_{E}(\eta) E^3
        + b_{E}(\eta) E^2 P + c_{E}(\eta) E P^2 + d_{E}(\eta) P^3 \right)
\ee
\be
  P' = - \sinh(\eta) E + \cosh(\eta)P + {1 \over {6 \k^2}} \left(a_{P}(\eta)
      E^3 + b_{P}(\eta) E^2 P + c_{P}(\eta) E P^2 + d_{P}(\eta) P^3 \right) .
\ee
For the special case $E = m_{0}$ and $P = 0$ we get:
\be
  \label{boosts}
  E' = \left( \gamma + {a_{E}(\eta) \over 6}({m_{0} \over \k})^2 \right) m_{0}
  \qquad
  P' = \left( \gamma + {a_{P}(\eta) \over{6 v}}
                         ({m_{0} \over \k})^2 \right) m_{0}v
\ee
where
\be
  v = \tanh(\eta)
  \qquad
  \gamma = \cosh(\eta) .
\ee
The functions $a_{E} , b_{E} , \dots d_{P}$ may be computed to any desired
order.

\Section{
         Conclusion
         }

At the energies near the Plank scale the structure of space-time is not
well-understood. Even at energies far below the Plank scale the validity
of \lor invariance is not clear \cite{Blo}.
If the structure of space-time at high energies is altered such that the
symmetry group of nature is affected, it may be replaced with a deformed
\poi group. This means that some deviations from
\lor transformations may be observed.
For the $\k$-deformed \poi group we find two classes of deviations.
Firstly, those which concern tachyonic states. As these do not
exist in nature, these deviations will not be verifiable. Secondly, there
are deviations for particles with real mass but at very high energies, or
in very fast moving frames. Here we mention two of such deviations.

Equations (\ref{tboosts}) mean that the spin degrees of freedom lag the
space-time degrees of freedom when boosting to high energies. This is because
the coefficient of $ K_i $ is less than the coefficient of $ L_i $, and $ K_i $
is the operator that boosts the spin degrees of freedom.
Also because of $ {1 \over{2 \k}} p S_{i} $ term in (\ref{tboosts}),
there is a rotation in spin degrees of freedom in such boosts. This may
results in an observable discrepancy in the spectra of atoms as seen from a
fast moving frame, or the spin flip of protons or electrons in an accelerator.

Referring to (\ref{boosts}) for an electron with energy one TeV we obtain:
\be
  P' \simeq E' = 10^6 ( 1 + 2.5 \times 10^{7} ({m_{0} \over \k })^2 ) m_{0}
\ee
where $ m_0 $ is electron's rest mass.
The discrepancy from Einstein's formula is
$2.5 \times 10^{7} ({m_{0} \over \k})^2$.
For a proton with the same energy the discrepancy is of the order
$10^5 ({M_{0} \over \k})^2$ where $M_{0}$ is the proton's rest mass.
Although proton is not a point particle, the trivial co-product of the
$\k$--deformed \poi group for energy $P_{0}$ justifies this reasoning.
\vskip 10 mm
{\bf Acknowledgments }
     \\
     This research was supported in part  by the
     Vice Chancellor for Research, Sharif University of
     Technology and in part by Institute for Studies in
     Theoretical Physics and Mathematics. The authors
     would like to thank A. Aghamohammadi, F. Ardalan,
     H. Arfaei, M. Khorrami and V. Karimipour for fruitful
     discussions.

\end{document}